\documentstyle{mn}
\input{psfig}

\title{NICMOS and VLBA observations of the gravitational lens system B1933+503}

\author[Marlow et al.]  
{D.~R. Marlow, I.~W.~A. Browne,
N. Jackson \& P.~N. Wilkinson \\
University of Manchester, NRAL Jodrell
Bank, Macclesfield, Cheshire SK11 9DL, England}

\date{August 20, 1998}

\begin{document}
\maketitle

\vskip 3mm
\begin{abstract}
NICMOS observations of the complex gravitational lens system B1933+503
reveal infrared counterparts to two of the inverted spectrum radio
images. The infrared images have arc-like structures. The corresponding
radio images are also detected in a VLBA map made at 1.7-GHz with a
resolution of 6~mas. We fail to detect two of the four inverted radio
spectrum components with the VLBA even though they are clearly visible
in a MERLIN map at the same frequency at a different epoch.  
The absence of these two
components could be due to rapid variability on a time-scale
less than the time delay, or to broadening of the images during
propagation of the radio waves through the ISM of the lensing galaxy
to an extent that they fall below the surface brightness detectability
threshold of the VLBA observations. The failure to detect the same two
images with NICMOS is probably due to extinction in the ISM of the
lensing galaxy.

\end{abstract}

 

\section{Introduction}

The Cosmic Lens All-Sky Survey (CLASS) is a survey of flat-spectrum
radio sources whose primary purpose is the discovery of new radio-loud
gravitational lens systems (Browne et al. 1997; 
Myers et al. 1999). A survey of
$\sim$15,000 radio sources has been made. The discovery of the complex
lens system B1933+503 has been reported by Sykes et al. (1998) and a
discussion of the lens model has been given in Nair (1998). The radio
maps (see Figure 1) show nine components, one of which (labelled 2)
probably consists of two blended images. The lens model indicates
that the unlensed radio source has an almost linear, three-component
structure on a scale of 200\--250 mas, with an
inverted\--spectrum component in the middle. Two of its components are
quadruply imaged and the third is doubly imaged. Evidence for flux
density variability at a wavelength of 2~cm is reported by Sykes et
al. (1998). The images corresponding to the central component of the source
are compact in MERLIN maps (labelled 1, 3, 4 and 6). 
Sykes et al. (1998) compare MERLIN and VLA maps
and find that the images of the central component have inverted radio
spectra. The relative flux densities of these inverted spectrum
components appear to have varied. The other images have steep spectra
and some are extended on the scale of $\sim$100~mas.

Sykes et al. (1998) also show an HST WFPC2 picture of the B1933+503
system taken through an I-band filter (F814W). The lensing
galaxy is clearly seen but no counterparts to any of the lensed images
are detected down to a limit of about magnitude 24.2 in I. A redshift
of 0.755 has been measured for the lensing galaxy (Sykes et al. 1998).

The presence of many images means that the lens mass model is well
constrained and the fact that some of the images are variable means
that it should be possible to measure time delays and hence determine
the Hubble constant. Because this is such a promising system we have
embarked on a systematic programme to obtain observational constraints
for the mass model.  In this paper we present HST NICMOS H\--band observations
that reveal for the first time infrared counterparts to two of
the radio images. We also present a high resolution VLBA 1.7-GHz map which
shows even more complex radio structure than has been seen before. We
discuss reasons why the flux density ratios of the images are very
different when measured with NICMOS and the VLBA to those measured
with MERLIN and the VLA.

\section{NICMOS  observations}

NICMOS observations of B1933+503 were made on October 9, 1997. The 
NIC1 camera was used providing a pixel scale  of $43~\rm{mas/pixel}$. 
A fit to the point spread function generated by the TinyTim 
program (Krist 1997) yields a FWHM of 131~mas for the NICMOS 
observations.
The F160W spectral filter was used which has a central wavelength 
of $1.6\mu\rm{m}$ and corresponds roughly to H\--band. 
The total integration time was 10495~seconds.

Both the lensing galaxy and two extended arc-like features (hereinafter
arcs) were detected. Parameters measured from the NICMOS picture are
listed in Table 1. The arcs correspond exactly in separation and
position angle to images 1 and 4 seen in the radio maps (see Figure
1). Moreover, the best-fit lens model described by Nair (1998) 
yields a lens centre (with respect to component 4) 
at (RA, Dec) $423^{+9}_{-5}, 270^{+7}_{-5}$ mas -- within $\sim$15~mas of 
the observed NICMOS separation of $417, 282 \pm 10$~mas.  

Thus there is no doubt that the arcs are
gravitational lens images of the host of the compact radio source. The fact
that arcs rather than point sources are detected indicates that
extended starlight dominates any emission from a compact active nucleus
in the lensed object.  Remarkably, no infrared emission is detected from the
positions of the other compact, inverted spectrum, radio components 3
and 6, indicating that the ratios of their flux densities with 
 component 1 must be $< 0.1$. In the VLA and MERLIN radio maps presented by
Sykes et al. (see their Table 2) the 3 to 1 and 6 to 1 ratios are
typically $\sim$0.7; i.e. very different from those in the
infrared. We discuss this discrepancy and a similar one in the
VLBA ratios in Section 4.

\section{VLBA 1.7-GHz observations}

The VLBA 1.7-GHz observations were made on June 25, 1996.  Phase
referencing was performed by alternately observing the strong
calibrator source B1954+51 and the target source 
over the total integration time of 15 hours. One VLA antenna
was included in the array to improve coverage of the $uv$ plane over
short spacings and to increase sensitivity to weak and extended
structure.  Fringe fitting was performed using B1954+51 and the
residual phase errors were corrected over the whole data set.  The data
reduction was performed in AIPS and mapping was performed using
DIFMAP, part of the Caltech VLBI package (Shepherd et al. 1994).

The naturally-weighted full-resolution VLBA map is shown in Figure 1. 
The resolution of the map is 6~mas and it has a noise level of $38\mu\rm{Jy}$. 
A separate map convolved with a 20~mas FWHM beam is also shown.
Individual component flux densities measured from the different 
maps are listed in Table 1.

There are several noteworthy results of these observations:
The images 2 and 5 are highly stretched into arc\--like features
consistent with the MERLIN map; 
images of the inverted spectrum components 1 and 4 are clearly
detected and component 4 has been resolved; 
there is no sign of the other inverted spectrum components 3 and
6 even though they are clearly present in the MERLIN L-band map
(Figure 1).

The 5-$\sigma$ detection limit on components 3 and 6 is roughly
an order of magnitude less than the 2.5~mJy and 2.2~mJy seen in the
MERLIN map. In order to search for some evidence of the missing images
a map was constructed from the VLBA data using only the shortest
(0--5M$\lambda$) baselines. In this map there were 2-$\sigma$ hints of
components 3 and 6 at the sub-mJy level suggesting perhaps that they
are highly resolved and have structure on the scale of $\sim$50~mas.

\section{Discussion}

The B1933+503 system is in many ways a promising one to use for the
determination of the Hubble constant from time delays because of its
reported variability and the multiple constraints on the mass
model. However, the lack of a redshift for, or even the detection of,
the lensed object has hitherto been a problem. The detection with
NICMOS of infrared emission from the lensed object now gives hope that
a redshift can be obtained. The fact that the NICMOS picture shows
extended arcs of emission rather than point sources suggests that we
are detecting the lensed emission from an intrinsically extended 
source. This is consistent with the faintness of the two images. Assuming no
extinction, if the lensed object has the luminosity of an elliptical L$^{*}$
galaxy, we would expect it to have approximately the observed magnitude of
H=22.1 for $z\la 2$ (Poggianti 1997). Since lensing also magnifies 
the images by about a magnitude the faintness of the objects implies 
one or more of the following: the lensed object is sub-L$^{*}$, it 
is $z \ga 2$, or it suffers extinction in the lensing galaxy.

The failure to detect images 3 and 6 in the NICMOS picture and in the
1.7-GHz VLBA maps (Figure 1) is intriguing. For the radio case we
consider two possibilities; either a combination of rapid variability
and lens time delay have conspired to move the images below the
detection limit of our VLBA observations, or they are of much lower
surface brightness (e.g. due to scattering) than the other images and
have been resolved out. We find neither explanation compelling.  The
latter seems unlikely, firstly because components 3 and 6 were
detected in September 1995 in 1~mas resolution, VLBA 5-GHz snapshot
observations (Sykes et al. 1998), (this result was consistent with
the expectation that inverted spectrum radio components should be
compact), and secondly, because the weaker, lower magnification, images
should be more compact than the stronger images.  However, it is
possible that the images could be scatter broadened by passage through
the ISM of the lensing galaxy. This is supported by the hint of a
detection of these images in the highly tapered VLBA 1.7-GHz map,
suggesting that they may have a much larger angular size than the
corresponding images 1 and 4. In another lens system, 0218+357, there
is a strong dependence of the image size on frequency of observation
(Patnaik and Porcas 1998) and it is quite likely that this arises from
interstellar scattering in the ISM of the lensing galaxy (Wilkinson et
al. 1998).

We must also consider the variability explanation for the
non-detection of images 3 and 6 in the VLBA map. There is already
evidence of relative variability amongst images 1, 3, 4 and 6 from VLA
2-cm observations (Sykes et al. 1998) of a few tens of percent. 
However, in extragalactic radio sources, flux density
variations at wavelengths around 18~cm are usually less than a few
tens of percent and occur on time-scales of months (Webber et al.
1980). This lack of variability is also well illustrated by a 
comparison of the flux densities of sources in the 1.4-GHz NVSS 
(Condon et al. 1998) and FIRST (Becker et al. 1995) catalogues. 
The observations for the catalogues were taken at 
least six months apart, but $<<1\%$ of sources have flux densities 
that differ by $>30\%$. In order for the explanation to work in B1933+503 the
variability must occur on a timescale shorter than the time
delays. Since the redshift of the lensed object is unknown the
predicted time delays have some uncertainty. However, Nair (1998)
suggests the delays with respect to component 1 should be approximately 8~days
for component 3, 7~days for component 4 and 9~days for component 6.
Thus the timescale of the variability must be shorter than a few
days. The required amplitude of the variability can be estimated from
comparing the flux densities in the MERLIN 18cm map (Figure 1) with the
detection limit on the present VLBA map. Components 3 and 6 need to
have decreased in flux density by nearly an order of magnitude in a
couple of days. We conclude that the explanation, though theoretically
possible, requires such extreme variability properties for B1933+503
as to render it highly unlikely.

It is perhaps significant that the same two images (3 and 6) are also
not detected in the NICMOS image. If variability were the explanation
for both the radio and infrared non-detections, we would have to
attribute the fact that the same two images are missing to coincidence
since radio and optical variability are hardly ever correlated and,
moreover, the NICMOS and VLBA observations are separated in time by
far longer than the expected time delays. Rapid variability of the
infrared emission also seems physically implausible given that the images
are extended and are therefore dominated by
starlight. It is, therefore, much more likely that in the infrared
images 3 and 6 are weaker relative to 1 and 4 than they are in the low
resolution radio maps due to infrared extinction. Supporting this view
is the fact that both 3 and 6 lie along the major axis of the lensing
galaxy where one might {\it a priori} expect there to be higher
extinction. The amount of extinction required is large 
 enough (A$_{\rm{V}}\sim$10) to suggest that there might also be ionized
gas present along the line of sight, maybe enough to give
rise to sufficient multi-path scattering to cause the images to be resolved
with the VLBA. We note that in B0218+357, the other lensed system for
which there is evidence of scatter broadening of the radio components,
there is also evidence for large extinction occurring in the lensing
galaxy (Xanthopoulos et al. 1998). In the case of B0218+357 it is 
known that the lens is a spiral galaxy. Perhaps this is also the 
case for B1933+503.

\section{Conclusions}

The presence of copious extended structure in our VLBA map means that
even more observational constraints are now available for the lens
mass model.

With the current data it is premature to reach a firm conclusion about
why the VLBA and infrared image flux ratios are so different from the
MERLIN and VLA ratios reported by Sykes et al. (1998). It seems very
plausible that extinction is responsible for the dimming of the images
in the NICMOS image. We also  favour the multi-path
scattering explanation for the non-detection of the corresponding
radio images. Clearly a radio map with better surface brightness
sensitivity should tell us the answer and we are planning to make
global VLBI observations for this purpose. On the other hand, if it
turns out that variability is the answer, then B1933+503 is indeed an
exciting object for time delay measurement and Hubble constant
determination.  VLA monitoring observations are in progress.

\section*{Acknowledgments}

This research used observations with the Hubble Space Telescope,
obtained at the Space Telescope Science Institute, which is operated
by Associated Universities for Research in Astronomy Inc. under NASA
contract NAS5-26555.  The VLBA is operated by Associated Universities
for Research in Astronomy Inc. on behalf of the National Science
Foundation.  MERLIN is operated as a National Facility by NRAL,
University of Manchester, on behalf of the UK Particle Physics \&
Astronomy Research Council.  This work is supported in part by the
European Commission, TMR Programme, Research Network Contract
ERBFMRXCT96-0034 ``CERES''. DRM is supported by a PPARC studentship.
We are grateful to P.~Helbig and L.~Koopmans for their
 assistance with this work.

\begin{figure*}
\caption{MERLIN, VLBA and NICMOS images of B1933+503.  Top left:
MERLIN 1.7-GHz map convolved with a 100~mas beam. Contours  are
plotted at $-2.5, 2.5, 5, 10, 20, 40, 80\%$ of the peak  flux density
of $16.6~\rm{mJy/beam}$. The individual components are numbered
1a\--8. The map is centred on RA 19 34 30.950 Dec +50 25 23.600 
(J2000). Top right: VLBA 1.7-GHz map naturally\--weighted map at
full\--resolution. Contours are plotted at $-1.5, 1.5, 3, 6, 12, 24, 48,
96\%$ of the peak flux density of $11.5~\rm{mJy/beam}$. 
The noise level in the map is $38\mu\rm{Jy}$. The map is centred on 
RA 19 34 30.932 Dec +50 25 23.502 (J2000).  
Bottom left: The VLBA 1.7-GHz map convolved with a
20~mas beam. Contours are  plotted at $-1.5, 1.5, 3, 6, 12, 24, 48,
96\%$ of the peak flux  density of $13.1~\rm{mJy/beam}$. The noise
level in the map  is $45\mu\rm{Jy}$. Bottom right: NICMOS
$1.6\mu\rm{m}$ image showing  the lensing galaxy and two of the lensed
components. The expected positions of the other lensed images relative to 
components 1 and 4 (as measured from the MERLIN 1.7-GHz map) 
are overlaid on to the NICMOS figure. 
 }
\begin{tabular}{cc}
\setlength{\unitlength}{1mm}
\begin{picture}(480,480)
\put(0,0){\includegraphics{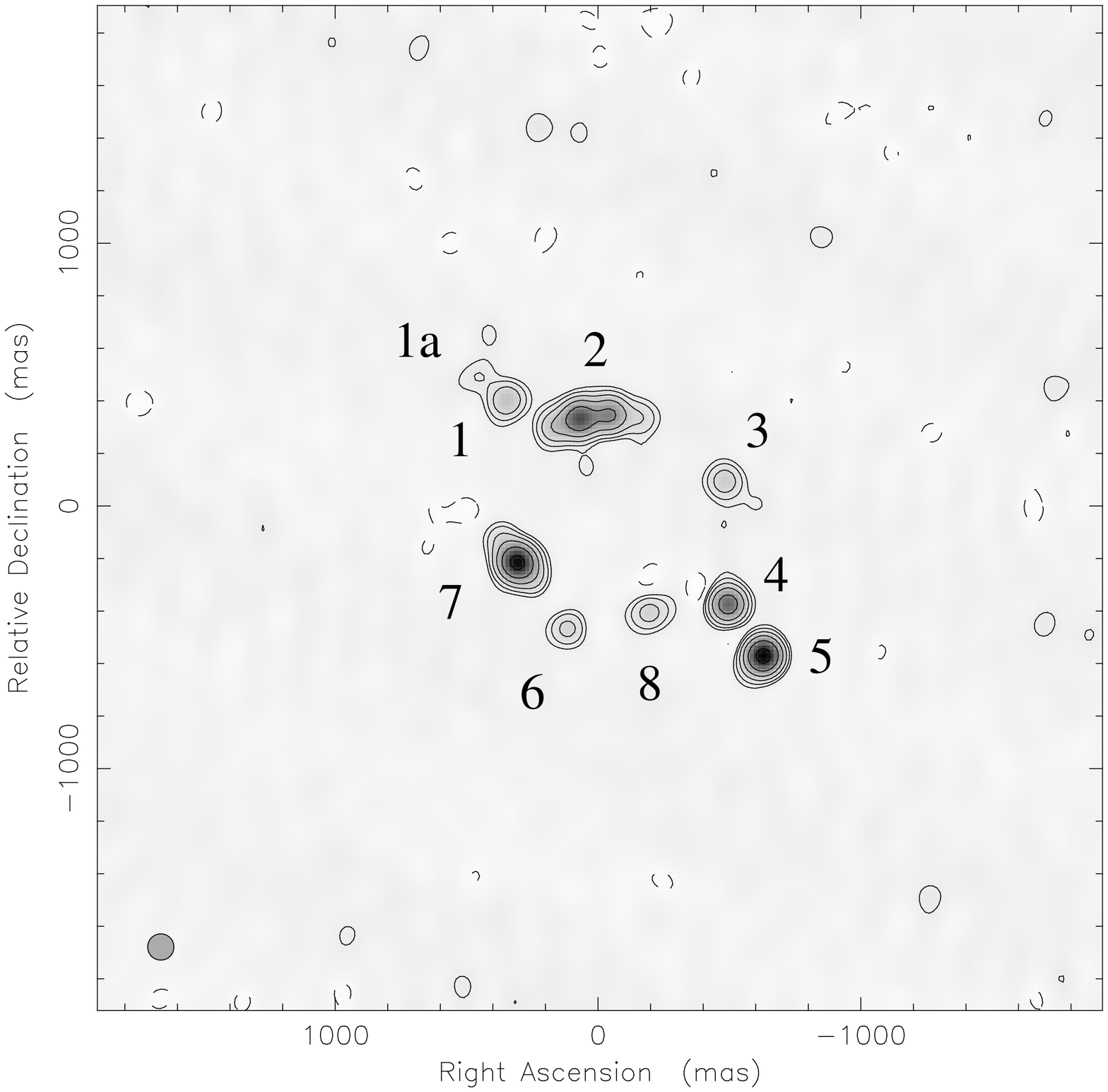}}
\put(100,0){\includegraphics{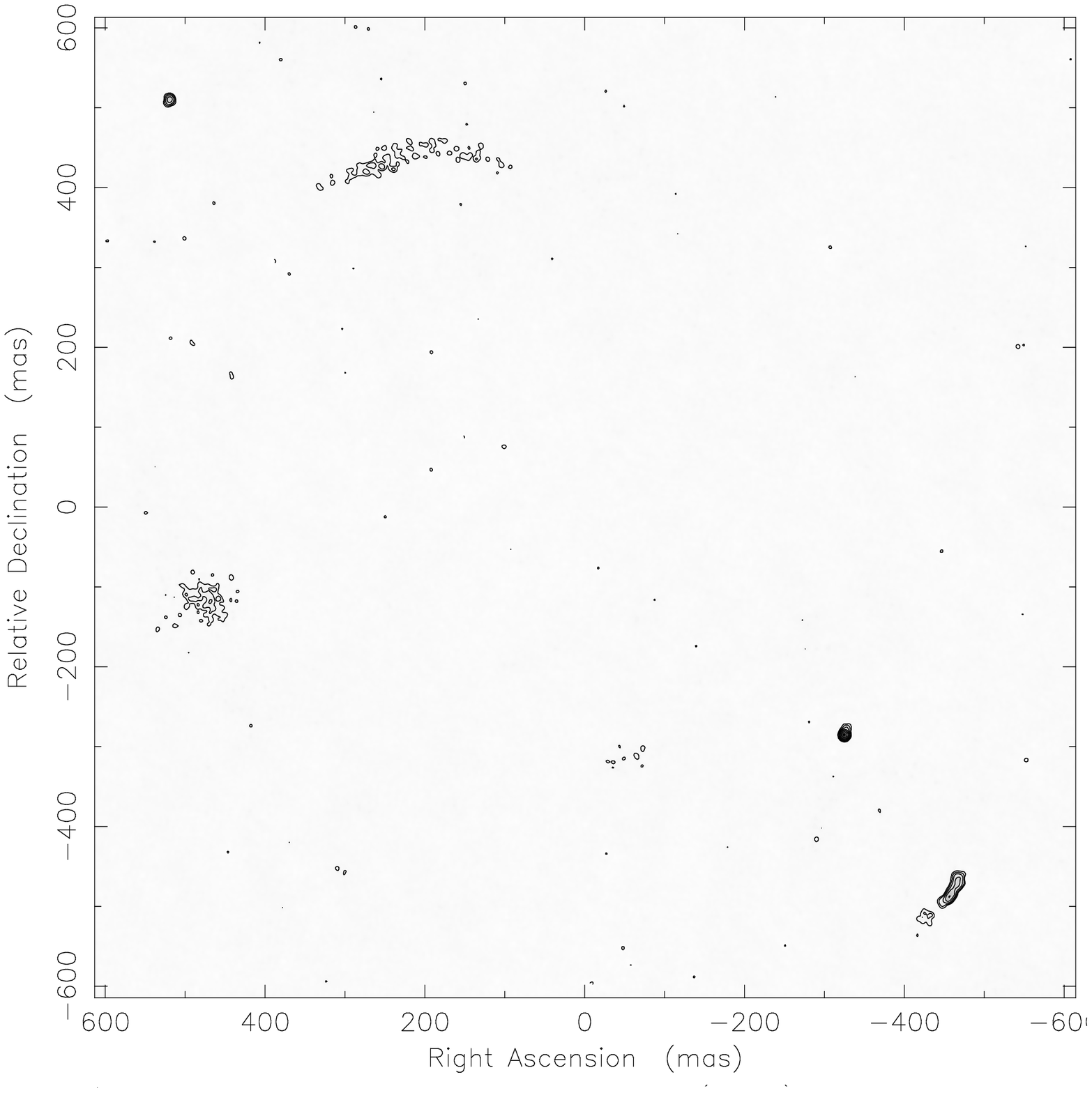}}
\put(0,100){\includegraphics{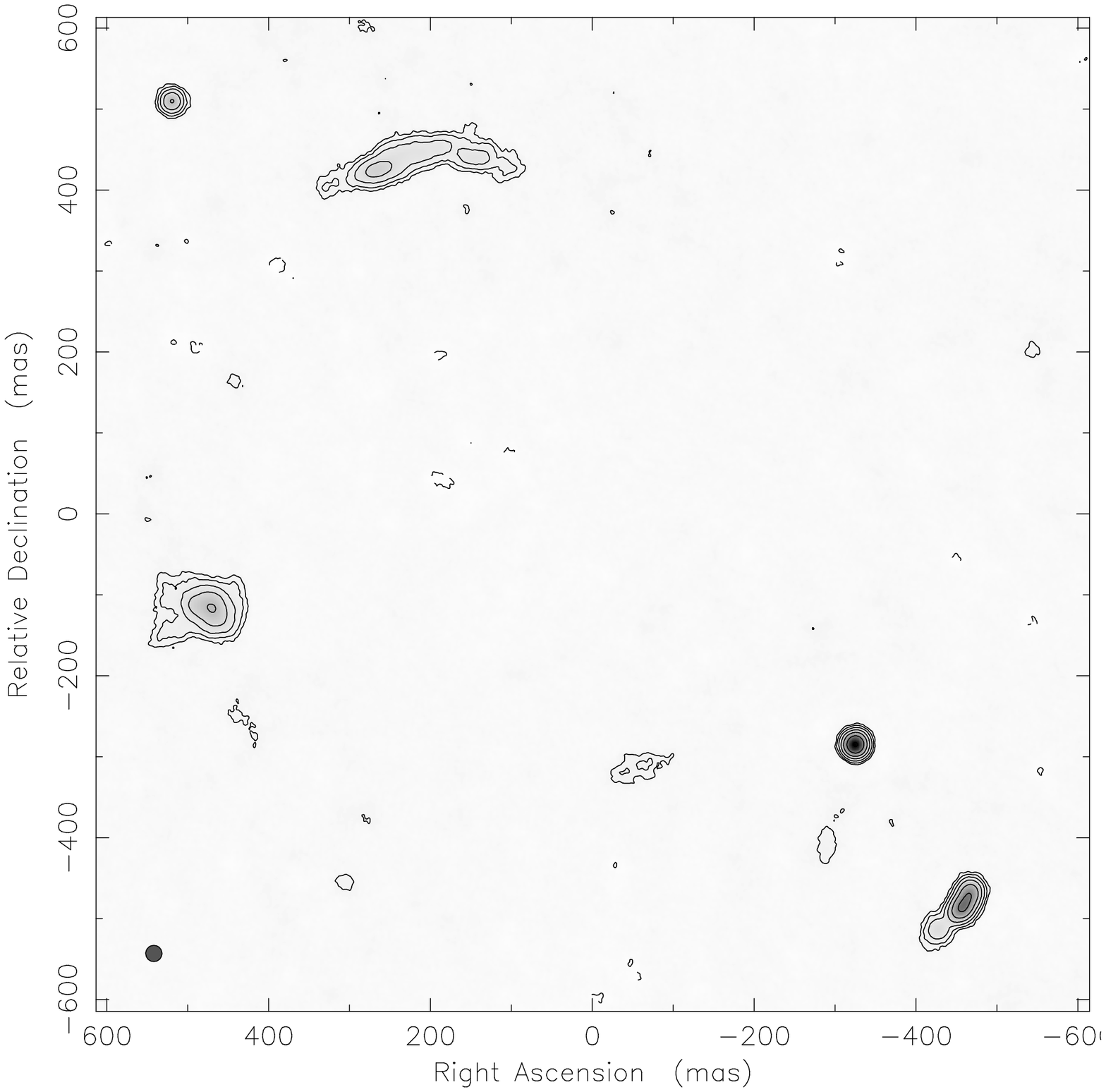}}
\put(100,100){\includegraphics{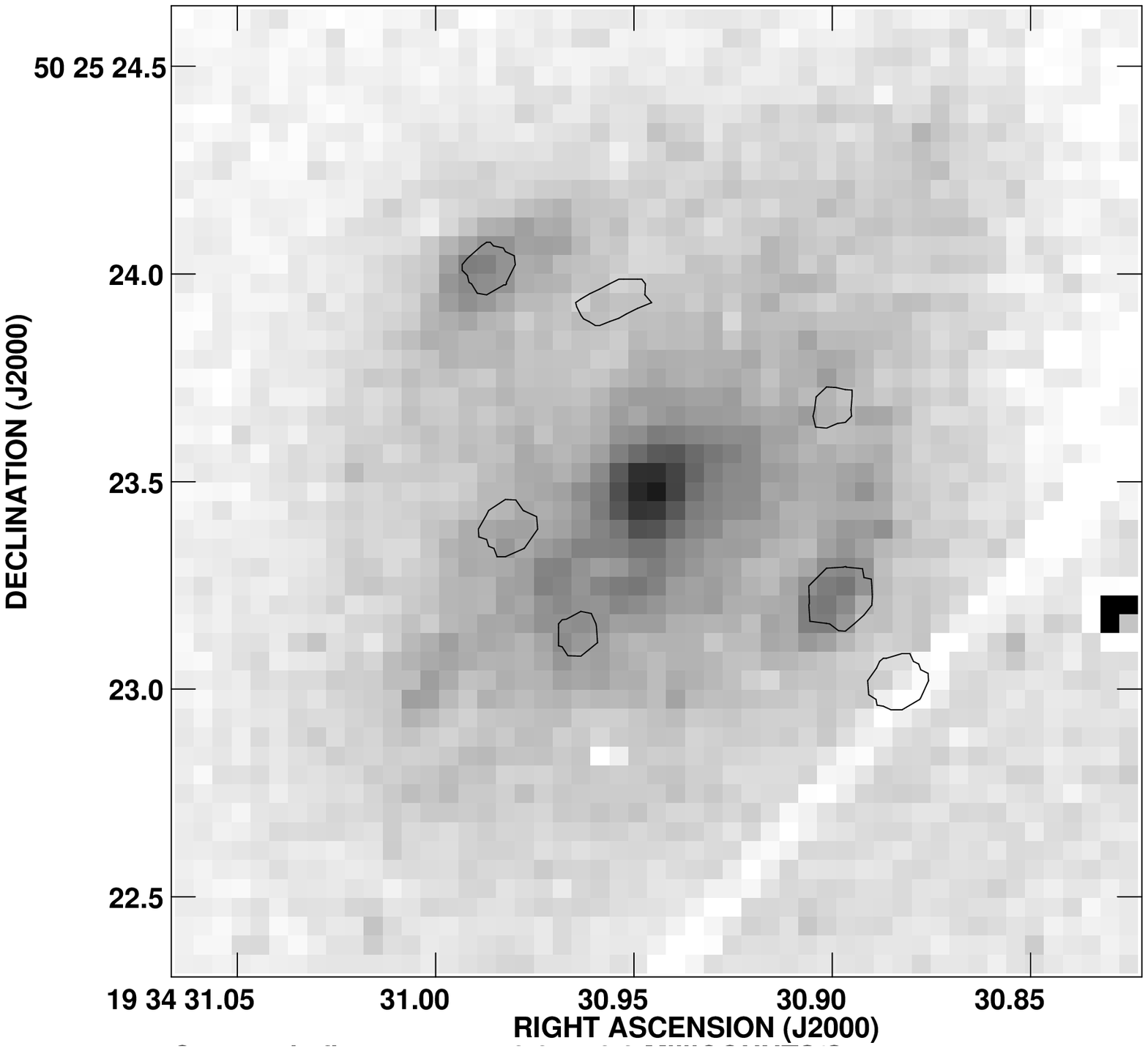}}
\end{picture}
\end{tabular}
\end{figure*}

\begin{table*}
\caption{Radio and F160W image flux densities and positions. Flux densities 
of components in the NICMOS image at $1.6\mu\rm{m}$ are given in the second 
column. The total integration time was 10495 seconds. Radio flux densities 
in various maps are quoted in columns 3, 4 and 5. Typical flux density errors 
are $\pm 0.4$~mJy. Values for  
the MERLIN observations of October 27, 1995 are taken from Sykes et al. 
(1998). Values are quoted for the VLBA observation 
in maps at the natural resolution of 6~mas and in one convolved with 
a 20~mas beam. Detection limits at $5\sigma$ are quoted for the missing 
components. Positions are given (in mas) 
for the the two lensed images 1 and 4 measured 
from the NICMOS image relative to the lensing galaxy centre.}
\begin{tabular}{cccccc}
\hline
Cmpt.& Flux density& Flux density& Flux density & Flux density &Offset from galaxy centre\\
&      $1.6\mu\rm{m}$&MERLIN 1.7-GHz &VLBA 1.7-GHz  
& VLBA 1.7-GHz & NICMOS \\

&  ($\times 10^{-20}\rm{Wm^{-2}nm^{-1}}$)& mJy/b (100~mas)& mJy/b (20~mas) 
& mJy/b (6~mas) &  (RA,Dec,error)\\
\hline

1a&  $ <0.02 $ & 0.9& $ <0.22 $& $ <0.19 $& -\\
1& $0.177\pm0.01$& 3.6& 3.5& 3.7& 406,538$\pm10$\\
2&    $ <0.02 $& 23.0&19.3& 35.9&-\\
3&    $ <0.02 $& 2.5& $<0.22 $& $ <0.19 $&-\\
4& $0.180\pm0.01$& 9.4&13.7&14.2&$-417$,$-282\pm10$\\
5&    $ <0.02 $& 16.2&18.3&18.7&-\\
6&    $ <0.02 $& 2.2& $<0.22 $& $<0.19 $&-\\
7&    $ <0.02 $& 20.3&18.2&20.4&-\\
8&    $ <0.02 $& 3.6&1.7&3.6&-\\
Gal&  $5.400\pm0.30$& -& - & -& 0,0\\

\hline
\end{tabular}

\end{table*}


\begin{thebibliography}{}
\bibitem{}
Becker, R.H., White, R.L. \& Helfand, D.J. 1995, ApJ, 450, 559
\bibitem{}
Browne, I.W.A., Jackson, N.J., Augusto, P., Henstock, D.R. Marlow, D.R., Nair, S. 
\& Wilkinson, P.N., 1997, Cosmology with the new radio surveys, 
eds Bremer, M., Jackson, N.J, I. P\'erez-Fournon (Kluwer Academic Publishers)
\bibitem{}
Condon, J.J., Cotton, W.D., Greisen, E.W., Yin, Q.F., Perley, R.A., Taylor, G.B., 
\& Broderick, J.J. 1998, AJ, 115, 1693)
\bibitem{}
Krist, J., 1997, ``The TinyTim user guide'', 
available at http://scivax.stsci.edu/~krist/tinytim.html
\bibitem{}
Myers, S., et al., 1999, in preparation
\bibitem{}
Nair, S. 1998, MNRAS, submitted, astro-ph/9803076.
\bibitem{}
Patnaik, A.R., Porcas, R.W. 1998, Highly redshifted radio lines, eds C. Carilli,
 S. Radford, K. Menten, G. Langston (PASP Conference series)
\bibitem{}
Pogianti, B.M., 1997, A\&AS, 122, 399 
\bibitem{}
Shepherd, M.C., Pearson,T.J. \& Taylor, G.B. 1994, BASS, 26, 987
\bibitem{}
Sykes, C.M. et al., 1998, MNRAS, submitted, astro-ph/9710358
\bibitem{}
Webber, J.C., Yang, J.S., Swenson, G.W. 1980, AJ, 85, 1434
\bibitem{}
Wilkinson, P.N. et al., 1998, in preparation
\bibitem{}
Xanthopoulos, A. et al., 1998, MNRAS, submitted
\end{thebibliography}
\end{document}